\documentclass[aps,prb,reprint,twocolumn,superscriptaddress,floatfix]{revtex4-1}

\usepackage{graphicx}  

\usepackage{amsmath,bm}   

\usepackage{dcolumn}

\begin{document}


\title{Intrinsic point defects and the $n$- and $p$-type dopability of the narrow gap semiconductors GaSb and InSb}

\author{J. Buckeridge} 
\email{j.buckeridge@ucl.ac.uk}
\affiliation{University College London, Department of Chemistry, 20 Gordon Street, London WC1H
  0AJ, United Kingdom}
\author{T. D. Veal}
\affiliation{Stephenson Institute for Renewable Energy and Department
  of Physics, School of Physical Sciences, University of Liverpool,
  Liverpool L69 7ZF, United Kingdom}
\author{C. R. A. Catlow}
\affiliation{University College London, Department of Chemistry, 20 Gordon Street, London WC1H
  0AJ, United Kingdom}
\author{D. O. Scanlon}
\affiliation{University College London, Department of Chemistry, 20 Gordon Street, London WC1H
  0AJ, United Kingdom}
\affiliation{Diamond Light Source Ltd., Diamond House, Harwell Science
  and Innovation Campus, Didcot, Oxfordshire OX11 0DE, United Kingdom}
\affiliation{Thomas Young Centre, University College London, Gower
  Street, London WC1E 6BT, United Kingdom}

\begin{abstract}


The presence of defects in the narrow-gap semiconductors GaSb and InSb affects their dopability and hence applicability for a range of optoelectronic applications. Here, we report hybrid density functional theory based calculations of the properties of intrinsic point defects in the two systems, including spin orbit coupling effects, which influence strongly their band structures. With the hybrid DFT approach we adopt, we obtain excellent agreement between our calculated band dispersions, structural, elastic and vibrational properties and available measurements. We compute point defect formation energies in both systems, finding that antisite disorder tends to dominate, apart from in GaSb under certain conditions, where cation vacancies can form in significant concentrations. Calculated self-consistent Fermi energies and equilibrium carrier and defect concentrations confirm the intrinsic $n$- and $p$-type behaviour of both materials under anion-rich and anion-poor conditions. Moreover, by computing the compensating defect concentrations due to the presence of ionised donors and acceptors, we explain the observed dopability of GaSb and InSb.

\end{abstract}

\maketitle
 

\section{Introduction}
\label{intro}

GaSb and InSb belong to the family of III-V, zinc blende structured semiconductors of interest from both a fundamental and technological point of view. The incorporation of Sb in III-V semiconducting nitrides, phosphides and arsenides results in a red shift of the band gap, opening up the possibility of pushing the frequency domain of devices based on such materials far into the infrared (IR).~\cite{gasb_expt_review_jap_dutta1997,long-wavelength_detectors_jap_rogalski2009,insb_expt_thz_transmission_quantum_oscillations_prl_gogoi2017} Both GaSb and InSb have applications in long wavelength telecommunications,~\cite{inassb_expt_growth_characterise_jvacscitechb_tomasulo2018} high speed microelectronics~\cite{insb_expt_highspeed_fet_apl_ashley1995,insb_expt_quantumwell_highmob_gasb_substrate_prm_lehner2018,insb_expt_diode_mid_ir_semicondscitech_petrosyan2019} and optoelectronics.~\cite{gasb_expt_vecsel_ingasb_qw_jcrystgrowth_paajaste2009,vecsels_review_jphysd_guina2017} Due to favourable lattice matching, GaSb can be used as a substrate for a wide range of ternary and quaternary III-V compounds.~\cite{gainsbn_expt_lattmatch_gasb_jcrysgrth_jefferson2007,gainsbn_expt_lattmatch_gasb_jap_ashwin2013,gasb_expt_ptype_acceptors_apl_kala2015,gasb_expt_epilaters_optical_semicondscitech_serincan2019} The spin-orbit interaction (SOI) has a strong effect on the valence band structure of both systems,~\cite{ii-v_ii-vi_expt_valence_bands_xps_prb_ley1974,gasb_inas_expt_heterostructure_topolins_disorder_prm_shojaei2018,gasb_expt_pdoped_effmass_qw_prb_karalic2019} but is more pronounced in InSb,~\cite{insb_expt_rashba_split_prb_khodaparast2004,insb_inas_nanowires_kp_soc_prb_campos2018} which, combined with a large Land\'{e} $g$-factor (over 50),~\cite{insb_expt_kp_lande_gfactor_prb_litvinenko2008} has meant that InSb has attracted considerable attention in the field of Majorana physics.~\cite{insb_expt_majorana_fermions_science_mourik2012,insb_expt_majorana_nanolett_deng2012} Moreover, GaSb and InSb have both been demonstrated to incorporate N and Bi effectively, resulting in a reduction in band gap~\cite{insbn_expt_growth_solstatelec_ashley2003,gasbn_expt_struct_opto_apl_veal2005,gasbn_expt_ftir_kp_apl_jefferson2006,gasbn_expt_mbe_growth_jcrysgrth_buckle2005,gasbn_gainsbn_expt_growth_spieproc_ashley2006,gasbn_lda_bandgap_bowing_apl_belabbes2006,gasbn-gasb_expt_QW_PL_jap_kent2007,gasbn_insbn_tb_bands_prb_lindsay2008,gasbn_expt_PL_jap_wang2009,gasbn_expt_growthrates_aipadv_ashwin2011,gasbn_gaasn_hse06_bands_prb_virkkala2012,gasbn_expt_optabs_3bac_apl_mudd2013,gasbn_expt_growth_lattconst_jphysdapplphys_aswhin2013,insbbi_expt_growth_bands_apl_rajpalke2014,gasbn_dft_lvms_prb_buckeridge2014,iii-v_bi_dft_electronicstates_semicondscitech_polak2015,insbn_expt_bandgap_dft_apl_linhart2016} in a similar manner to the more widely studied, GaAs-based dilute nitrides and bismides.~\cite{dilute-nitrides_review_semicondscitech_oreilly2009,iii-v_review_expt_growth_doping_applphysrev_kuech2016} Alloys can be produced of GaAs, GaSb and InSb, together with the relevant nitrides and/or bismides to tune the optical and electronic properties for a variety of applications;~\cite{ingaassbn_gaas_expt_alignment_apl_chang2014,iii-v_bi_k.p_qw_models_jap_gladysiewicz2016,insbnbi_calc_kp_topologicalins_qsh_newjphys_song2017,inassbbi_expt_growth_bandgap_apl_webster2017,iii-v_dilute_n_model_solar_cell_semicondscitech_kharel2019} indeed, very high efficiency tandem solar cells include an active layer composed of such an alloy.~\cite{dilute-nitride_world-record_solarcell_mrsproc_jones-albertus2013} 



Given the importance of GaSb and InSb, there are surprisingly few studies on their intrinsic defect properties, which are key to their dopability and hence functionality in devices. As-grown GaSb has been shown to be $p$-type regardless of growth conditions,~\cite{gasb_expt_ptype_hall_physrev_leifer1954,gasb_expt_crystal_growth_hole_conc_jphyschemsol_vandermeulen1967,gasb_expt_ptype_undoped_semicondscitech_haywood1988,insb_expt_gasb_defects_compensate_intjhighspeedelecsys_pino2004,gasb_expt_ptype_acceptors_apl_kala2015,gasb_expt_pdoped_effmass_qw_prb_karalic2019} although the acceptor concentrations can be decreased slightly by varying the V/III flux when growing with molecular beam epitaxy (MBE).~\cite{gasb_expt_sdoped_mbe_compensation_jap_lee1990,gasb_expt_tedoped_highmob_jvacscitechb_turner1993} Gallium vacancies ($V_{\mathrm{Ga}}$) have been shown to occur in GaSb using positron annihilation spectroscopy (PAS),~\cite{gasb_expt_positron_hall_shallow_acceptor_apl_ling2004} but have been ruled out as the dominant acceptor; instead, it has been inferred in further PAS studies that the gallium antisite (Ga$_{\mathrm{Sb}}$) is responsible for the observed $p$-type activity,~\cite{gasb_expt_positron_point-defects_jap_kujala2014,gasbn_expt_dft_positron_defects_jap_segercrantz2015} based on earlier density functional theory (DFT) calculations using the local density approximation (LDA).~\cite{gasb_lda_intrinsic_defects_jap_hakala2002} While the LDA was also used to investigate the r\^{o}le of H in GaSb,~\cite{gasb_lda_so_h_defect_prb_peles2008} this approach suffers from the well-known band gap underestimation error, which is particulary problematic in narrow gap semiconductors such as GaSb and InSb. To overcome the band gap error, a subsequent study on defects in GaSb employed hybrid DFT (without including the SOI).~\cite{gasb_dft_hybrid_intrinsic_defects_prb_virkkala2012} The results, however, indicated that the intrinsic defect physics would result in a semi-insulating material as-grown, in contrast to experiment. C and O impurities were instead proposed to account for the $p$-type activity.

There are even fewer studies of the defect properties of InSb. The material can be made $n$- or $p$-type depending on growth conditions, while temperature ($T$) dependent studies have been employed to study variations in the $n$-type carrier concentration, Fermi energy and mobilities in order to elucidate various defect properties.~\cite{insb_expt_hall_conductivity_effmass_physrev_hrostowski1955,insb_expt_fermi_energy_t_pssb_zukotynski1970,insb_expt_intrinsic_carrierconc_jap_chen1972,insb_expt_carrierconc_effmass_vs_t_hall_jphyschemsol_oszwaldowski1988,insb_expt_gasb_defects_compensate_intjhighspeedelecsys_pino2004,insb_expt_growth_sb_antisite_jcrystgrowth_jin2011} A computational study using DFT with the LDA indicated that the antimony antisite (Sb$_{\mathrm{In}}$) would dominate in Sb-rich growth conditions;~\cite{insb_dft_inp_inas_lda_bulk_110_defects_prb_hoglund2006} by varying growth conditions, it was suggested that the formation of this defect could be suppressed in epitaxially grown thin films.~\cite{insb_expt_growth_sb_antisite_jcrystgrowth_jin2011} Furthermore, it has been proposed that the formation of indium vacancies as well as Sb$_{\mathrm{In}}$ can account for observed changes in the electronic properties of InSb grown in varying conditions.~\cite{insb_expt_mbe_growth_chinphyslett_zhao2017} To our knowledge, no comprehensive study on the intrinsic defects in InSb using hybrid DFT has yet been performed.

In this Paper, we use hybrid DFT, including the SOI, to investigate the dominant native point defects in both GaSb and InSb. As noted above, the SOI strongly affects the dispersion of the upper valence bands in both systems; therefore, depending on the composition of the particular defect states, it can have a significant effect on the defect formation energies. We tune the fraction of exact exchange in the hybrid functional to reproduce only the band gaps, and justify this approach by computing a range of bulk properties of both systems, demonstrating close agreement with experiment for the structural, electronic, elastic and lattice vibrational properties. Our results show that GaSb will be $p$-type when grown in Sb-poor conditions, but may be semi-insulating under Sb-rich conditions. InSb, in contrast, will be $n$-type under Sb-poor conditions and $p$-type under Sb-rich conditions. From our computed defect formation energies, we determine self consistent Fermi energies and equilibrium carrier and defect concentrations as a function of $T$, by imposing the constraint of charge neutrality, calculating concentrations that agree well with experiment. Moreover, by introducing fixed concentrations of fully ionised dopants into the self-consistent Fermi energy calculation, we investigate donor and acceptor compensation by native defects in both systems. We find that, while InSb can be easily $n$- or $p$-doped, GaSb cannot be effectively $n$-doped under Sb-poor conditions. We provide the first comprehensive study of intrinsic disorder in GaSb and InSb using relativstic hybrid DFT which helps to elucidate the defect properties and dopability of both systems under equilibrium conditions. 

The rest of the paper is structured as follows: In Section~\ref{calc}, we describe our computationaly methodology. We present our results in Section~\ref{res} and summarize our main findings in Section~\ref{conclude}.

\section{Calculations}
\label{calc}

To calculate the bulk and defect properties of GaSb and InSb, we have used plane-wave DFT as implemented in the \texttt{VASP} code,~\cite{vasp_prb_kresse1993, vasp_prb_kresse1994, vasp_compmatsci_Kresse1996, vasp_prb_kresse1996} utilizing the Heyd-Scuseria-Ehrnzerof (HSE06) hybrid density functional~\cite{hse06_functional_jphyschem_heyd2006} for electron exchange and correlation with the projector augmented wave method~\cite{paw_physrevb50_blochl1994} to model the interaction between core and valence electrons (including $3d$ and $4d$ states among the 13 valence electrons in the cases of Ga and In, respectively, and five valence electrons for As). Spin-orbit interactions were included in all calculations.~\cite{spin-orbit_vasp_prb_hobbs2000} The proportion $\alpha$ of exact exchange in the hybrid functional was set to $\alpha=0.335$ ($\alpha=0.31$) for GaSb (InSb) in order to reproduce the fundamental gap (see below). The total energy of the zinc blende primitive cell was calculated at a series of constant volumes, using a 400\,eV plane wave cut off and a 12$\times$12$\times$12 $\Gamma$-centred Monkhorst-Pack~\cite{monkhorst_pack_prb_monkhorst1976} \textit{k}-point mesh (a finer 14$\times$14$\times$14 \textit{k}-point grid was used when computing the density of states (DOS)), which provided convergence in the total energy up to 10$^{-4}$ eV, fitting the resultant energy-volume data to the Murnaghan equation of state. The bulk modulus $B_0$ was derived using this approach. The zone-centre longitudinal phonon frequencies ($\omega_{\mathrm{LO}}$) were calculated using the frozen phonon approach, as implemented in \texttt{VASP}.~\cite{optical-prop_paw_rpa_prb_gajdos2006} We have also computed the elastic constants C$_{11}$, C$_{12}$ and C$_{44}$, using the finite displacement approach available in \texttt{VASP}. Electron ($m^*_{\mathrm{e}}$), light hole ($m^*_{\mathrm{lh}}$) and heavy hole ($m^*_{\mathrm{hh}}$) effective masses were calculated by fitting quadratic functions to the energy dispersion within 1 meV of the appropriate band extremum. For the hole masses, derived from the valence bands where the dispersion is non-spherical, we took an average of the values obtained for the different cartesian directions. 

Defect calculations were performed using the supercell approach with a 64-atom $2\times2\times2$ expansion of the conventional cubic cell, which has been shown to be suitably converged previously.~\cite{gasbn_dft_lvms_prb_buckeridge2014,gasb_lda_so_h_defect_prb_peles2008,gasb_dft_hybrid_intrinsic_defects_prb_virkkala2012,my_strain_paper,gaasn_hetdevice_mob_prb_buckeridge2011,gaasn_si_vib_modes_solstatcomm_buckeridge2010} The formation energy of defect X in charge state $q$, $E_f(\mathrm{X}^q)$, was determined through calculation of the heat of formation of the relevant defect reaction:~\cite{gaas_lda_nativedefects_prl_zhang1991,defect-calcs_review_revmodphys_freysoldt2014}
\begin{eqnarray}
E_f(\mathrm{X}^q) =&& E_{tot}(\mathrm{X}^q) - E_{tot}(\mathrm{bulk}) - \sum_i n_i \mu_i \nonumber \\*
&& + q(E_{\mathrm{VBM}} + \Delta + E_F) + E_{\mathrm{c}},
\end{eqnarray}
where $E_{tot}(\mathrm{X}^q)$ ($E_{tot}(\mathrm{bulk})$) is the total energy of the defect-containing (pure bulk) supercell, $E_{\mathrm{VBM}}$ is the energy at the valence band maximum (VBM), $E_F$ is the Fermi energy (introduced as a parameter), $\Delta$ is the energy required to align the electrostatic potential in the defect supercell with that of bulk and $E_{\mathrm{c}}$ is a correction term to account for supercell errors such as image charge interactions and, where applicable, erroneous band filling by delocalised carriers. To calculate $\Delta$ and $E_{\mathrm{c}}$, we follow the procedure outlined by Lany \textit{et al.},~\cite{lany-zunger_correction_prb_lany2008} which has been shown to result in corrections closely matched to those derived from full solutions to Poisson's equation.~\cite{charge_correction_poisson_solve_jchemphys_durrant2018} $n_i$ is the number of species $i$ that is added to ($n_i > 0$) or removed from ($n_i < 0$) the supercell to form $X$, and $\mu_i$ is the chemical potential of species $i$, taken with reference to the calculated standard state energies $E_i$ so that $\mu_i=E_i+\Delta\mu_i$.~\cite{cplap_cpc_buckeridge2014} The values of $\Delta\mu_i$ can vary depending on the environmental conditions in thermodynamic equilibrium, but are contstrained by the relation $\Delta\mu_{\mathrm{M}} + \Delta\mu_{\mathrm{Sb}} = \Delta H[\mathrm{MSb}]$, where M$=$Ga or In and $\Delta H[\mathrm{MSb}]$ is the heat of formation of MSb; we calculate $\Delta H[\mathrm{GaSb}]=-0.507$ eV and $\Delta H[\mathrm{InSb}]=-0.470$ eV, which are in reasonable agreement with the experimental values of -0.433 eV and -0.316 eV, respectively,~\cite{crc_handbook_89th_2008} particularly taking into account that the experimental values correspond to room $T$, while the calculations are done at the athermal limit (one would expect the heats of formation to become more negative by $\sim 0.05$ eV~\cite{crc_handbook_89th_2008} at 0 K).~\footnote{If we were to use the experimental heats of formation, there would be no significant difference in our conclusions.} We calculate the $E_f[\mathrm{X}]$ at two extremes: Sb rich, where $\Delta\mu_{\mathrm{Sb}}=0$ eV, corresponding to an excess of Sb in the growth environment and absence of pure In, and Sb poor, the opposite extreme, where $\Delta\mu_{\mathrm{Sb}}=\Delta H[\mathrm{MSb}]$.

From the calculated defect formation energies and DOS, we used the code \texttt{SC-FERMI}~\cite{scfermigithub,lafeo3_dft_defects_chemmater_taylor2016,taas_dft_nonstoich_prb_buckeridge2016,sc-fermi_accepted_compphyscom_buckeridge2019} to determine the equilibrium carrier and defect concentrations. \texttt{SC-FERMI} employs Fermi-Dirac statistics to calculate the concentrations, which are functions of $E_F$. With the constraint of overall charge neutrality in the system, a self-consistent $E_F$ can be derived at any temperature and consequently so can the electron ($n_0$), hole ($p_0$) and defect ($[\mathrm{X}]$) concentrations. Moreover, the charge neutrality constraint can be exploited in order to introduce fixed concentrations of ionised impurities, and the equilibrium carrier and defect concentrations recalculated in the presence of such impurities. In such a way, one can analyse ionised donor and acceptor compensation. In our calculations we neglect the temperature dependence of the free energies of defect formation due to the high computational cost in determining the associated vibrational entropy; one would expect the free energies to change by $\sim0.1-0.2$ eV over the temperature range we employ, but including such changes would not affect significantly the conclusions we draw from our results.

\section{Results}
\label{res}
\subsection{Bulk properties}
\label{bulk_props}

\begin{table*}[ht]
\caption{Calculated lattice parameter $a$, bulk modulus $B_0$, elastic constants C$_{11}$, C$_{12}$ and C$_{44}$, band gap $E_g$, spin-orbit split off energy $\Delta_{\mathrm{SO}}$, electron ($m^*_{\mathrm{e}}$), light hole ($m^*_{\mathrm{lh}}$) and heavy hole ($m^*_{\mathrm{hh}}$) effective masses and zone-centre longitudinal optical phonon frequency $\omega_{\mathrm{LO}}$ of GaSb and InSb, compared with experimental results.~\cite{gasb_expt_lowT_lattconst_doklakadsssr_siroto1962,gasb_expt_elastic_bulk_mod_pressure_jap_mcskimin1968,gasb_gap_expt_elastic_consts_prb_boyle1975,gasb_expt_lowt_bandgap_jap_wu1992,gasb_expt_bands_prb_chiang1980,gasb_expt_elec_eff_mass_jphyschemsol_hill1974,gasb_expt_hole_eff_mass_jap_heller1985,gasb_expt_raman_pressure_prb_cardona1984,iii-v_review_bandfeatures_jap_vurgaftman2001,insb_expt_elastic_const_physrev_slutsky1959,insb_expt_bandgap_t_apl_litter1985,insb_expt_hall_conductivity_effmass_physrev_hrostowski1955,small_gap_transport_advphys_zawadzki1974,insb_expt_phonons_prb_price1971} The effective masses are given in units of the electronic rest mass.} \centering
\begin{ruledtabular}
\begin{tabular} { c c | c c c c c c c c c c c }
& & $a$ (\AA\,) & $B_0$ (GPa) & C$_{11}$ (GPa) & C$_{12}$ (GPa) & C$_{44}$ (GPa) & $E_g$ (eV) & $\Delta_{\mathrm{SO}}$ (eV) & $m^*_{\mathrm{e}}$ & $m^*_{\mathrm{lh}}$ & $m^*_{\mathrm{hh}}$ &  $\omega_{\mathrm{LO}}$ (cm$^{-1}$) \\ \hline
GaSb & Calc. & 6.137 & 55.1 & 92.33 & 39.03 & 45.99 & 0.808 & 0.76 & 0.041 & 0.047 & 0.23 & 230.4 \\
& Expt. & 6.09593~\cite{gasb_expt_lowT_lattconst_doklakadsssr_siroto1962} & 56.35~\cite{gasb_expt_elastic_bulk_mod_pressure_jap_mcskimin1968} & 90.82~\cite{gasb_gap_expt_elastic_consts_prb_boyle1975} & 41.31~\cite{gasb_gap_expt_elastic_consts_prb_boyle1975} & 44.47~\cite{gasb_gap_expt_elastic_consts_prb_boyle1975} & 0.813~\cite{gasb_expt_lowt_bandgap_jap_wu1992} & 0.82~\cite{gasb_expt_bands_prb_chiang1980} & 0.0412~\cite{gasb_expt_elec_eff_mass_jphyschemsol_hill1974} & 0.05~\cite{gasb_expt_hole_eff_mass_jap_heller1985} & 0.28~\cite{gasb_expt_hole_eff_mass_jap_heller1985} & 232.6~\cite{gasb_expt_raman_pressure_prb_cardona1984} \\ \hline
InSb & Calc. & 6.548 & 40 & 68.2 & 33.8 & 31.6 & 0.23 & 0.80 & 0.018 & 0.019 & 0.25 & 180.3 \\
& Expt. & 6.4794~\cite{iii-v_review_bandfeatures_jap_vurgaftman2001} & 48.1~\cite{insb_expt_elastic_const_physrev_slutsky1959} & 69.18~\cite{insb_expt_elastic_const_physrev_slutsky1959} & 37.88~\cite{insb_expt_elastic_const_physrev_slutsky1959} & 31.32~\cite{insb_expt_elastic_const_physrev_slutsky1959} & 0.24~\cite{insb_expt_bandgap_t_apl_litter1985} & 0.80~\cite{insb_expt_bandgap_t_apl_litter1985} & 0.015~\cite{insb_expt_hall_conductivity_effmass_physrev_hrostowski1955} & 0.015~\cite{small_gap_transport_advphys_zawadzki1974} & 0.43~\cite{small_gap_transport_advphys_zawadzki1974} & 196.8~\cite{insb_expt_phonons_prb_price1971} \\
\end{tabular}
\end{ruledtabular}
\label{bulk}
\end{table*}

In Table~\ref{bulk}, we show our calculated lattice parameter $a$, $B_0$, elastic constants C$_{11}$, C$_{12}$ and C$_{44}$, band gap $E_g$, spin-orbit split off energy $\Delta_{\mathrm{SO}}$, $m^*_{\mathrm{e}}$, $m^*_{\mathrm{lh}}$, $m^*_{\mathrm{hh}}$ and $\omega_{\mathrm{LO}}$ for GaSb and InSb, compared with experiment.~\cite{gasb_expt_lowT_lattconst_doklakadsssr_siroto1962,gasb_expt_elastic_bulk_mod_pressure_jap_mcskimin1968,gasb_gap_expt_elastic_consts_prb_boyle1975,gasb_expt_lowt_bandgap_jap_wu1992,gasb_expt_bands_prb_chiang1980,gasb_expt_elec_eff_mass_jphyschemsol_hill1974,gasb_expt_hole_eff_mass_jap_heller1985,gasb_expt_raman_pressure_prb_cardona1984,iii-v_review_bandfeatures_jap_vurgaftman2001,insb_expt_elastic_const_physrev_slutsky1959,insb_expt_bandgap_t_apl_litter1985,insb_expt_hall_conductivity_effmass_physrev_hrostowski1955,small_gap_transport_advphys_zawadzki1974,insb_expt_phonons_prb_price1971} As described above, the $\alpha$ used in the hybrid functional was chosen to reproduce the band gap at low $T$. From Table~\ref{bulk}, however, we see that the hybrid DFT approach reproduces very well the experimental structural, elastic, and lattice vibrational properties of both materials, while the energy dispersion derived properties are also well reproduced. The only significant discrepancies occur for InSb, particularly in $B_0$ and $\omega_{\mathrm{LO}}$, indicating a slightly softer lattice in the calculation compared with experiment. The calculated $m^*_{\mathrm{hh}}$ for InSb is significantly lower than the experimental value, but this discrepancy may be due to difficulties in measuring this property accurately. Overall, the agreement between the calculated values and experiment is satisfactory, and indicates that our DFT approach is appropriate. 

\begin{figure}[ht!]
\centering
\vspace{0.5cm}
\includegraphics*[width=1.0\linewidth]{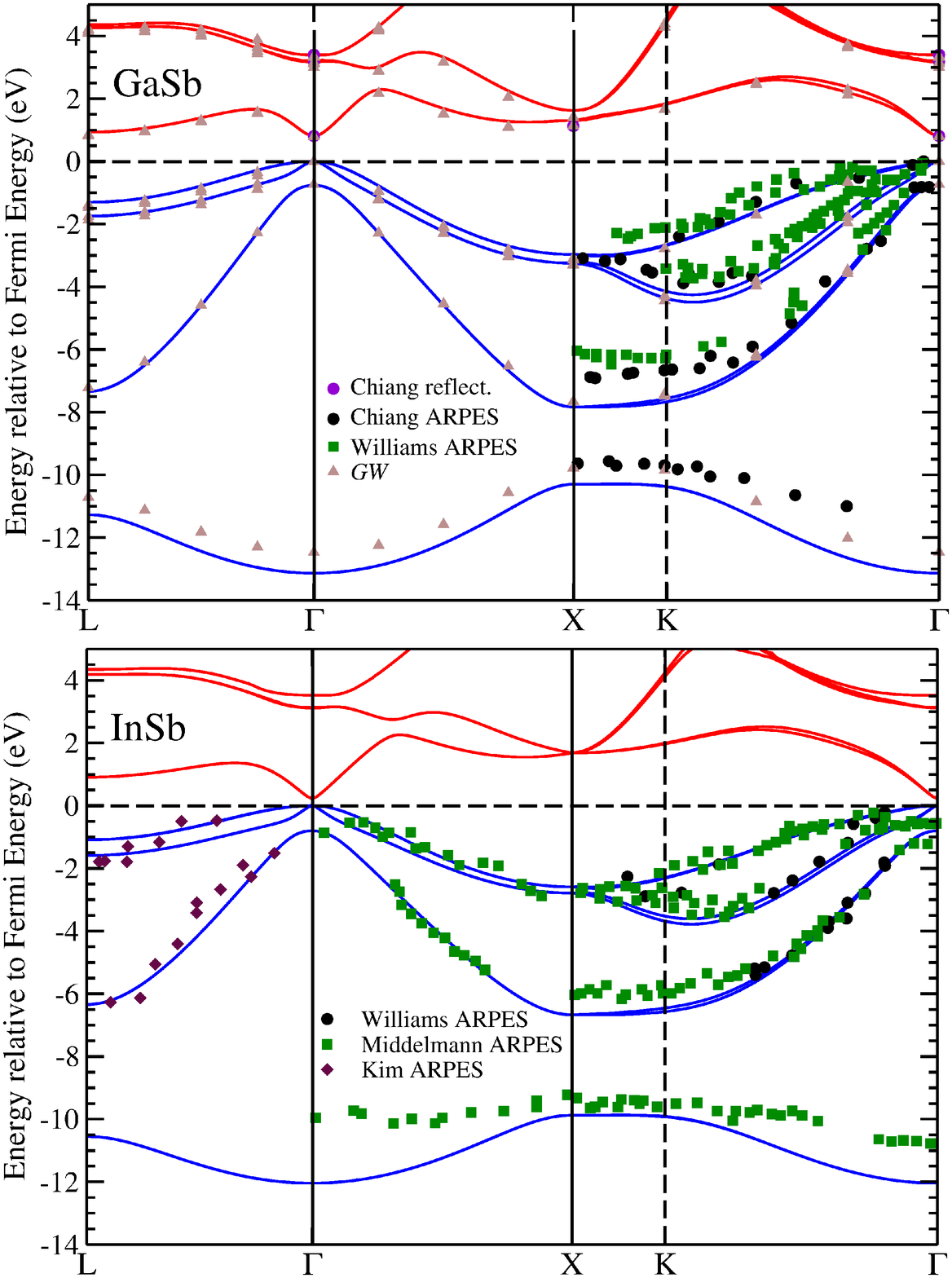}
\caption{(Color online) Band structure of GaSb and InSb calculated using hybrid density functional theory (valence bands indicated by blue lines, conduction bands by red lines), compared with experimental results determined for the case of GaSb using reflectance measurements by Chiang and Eastman~\cite{gasb_expt_bands_prb_chiang1980} (purple circles) and angle-resolved photoemission spectroscopy (ARPES, black circles and green squares) by Chiang and Eastman~\cite{gasb_expt_bands_prb_chiang1980} and Williams \textit{et al.}~\cite{gasb_insb_etc_expt_valencebands_arpes_prb_williams1986}, as well as calculated energy levels using self-consistent $GW$ (brown triangles). The InSb bands are compared with ARPES measurements by Williams \textit{et al.}~\cite{gasb_insb_etc_expt_valencebands_arpes_prb_williams1986} (black circles), Middelmann \textit{et al.}~\cite{insb_expt_arpes_synch_prb_middelmann1986} (green squares) and Kim \textit{et al.}~\cite{insb_arpes_111_jphyscondmat_kim1996} (maroon diamonds).}
\label{fig-bands}
\end{figure}

In Fig.~\ref{fig-bands}, we show our hybrid-DFT-computed band structures of GaSb and InSb compared with experimental values determined using angle-resolved photoemission spectroscopy (ARPES) and, for the case of GaSb, reflectance measurements.~\cite{gasb_expt_bands_prb_chiang1980,gasb_insb_etc_expt_valencebands_arpes_prb_williams1986,insb_expt_arpes_synch_prb_middelmann1986,insb_arpes_111_jphyscondmat_kim1996} For GaSb, we have also calculated band energies using the fully self consistent $GW$ approach, as implemented in \texttt{VASP},~\cite{vasp_gw_paw_prb_shishkin2006,vasp_gw_selfconsis_prb_shishkin2007,vasp_gw_vertex_prl_shishkin2007} including the SOI. As these calculations are computationally expensive, we have not determined the dispersion along the high symmetry path in the Brillouin zone with as small a grid spacing as we have for the hybrid DFT calculations. The band structure is similar in both cases to GaAs,~\cite{yuandcardona2} with the VBM and conduction band minimum (CBM) both occuring at the $\Gamma$ point, and a splitting of the 6-fold degenerate upper valence bands into 4-fold and 2-fold degenerate bands, the latter forming the spin-orbit split-off bands. For both systems, the hybrid DFT approach reproduces the band structure well, apart from the lower-lying Sb s states (at about -11 eV), which are deeper than either experiment or the $GW$ results. The bands near the VBM and the conduction band minimum (CBM), however, are very well reproduced. These bands are the most significant for defect state formation. 


\subsection{Defects in GaSb}
\label{gasb}

\begin{figure}[ht!]
\centering
\vspace{0.5cm}
\includegraphics*[width=1.0\linewidth]{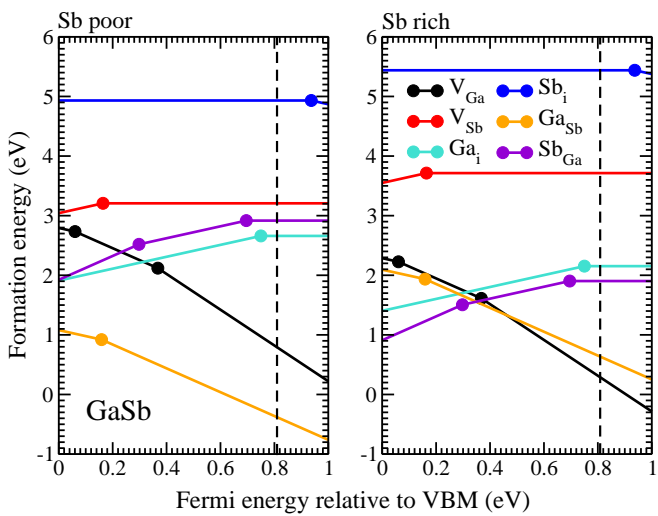}
\caption{(Color online) Calculated formation energies of each intrinsic defect (vacancies, interstitials and antisites; see text for description) in GaSb as a function of Fermi energy relative to the valence band maximum (VBM), shown for Sb-poor and Sb-rich conditions. The slope of each line indicates the defect charge state; the transition levels lie where the slopes change. The dashed line indicates the position of the conduction band minimum.}
\label{fig-gform}
\end{figure}

Our calculated formation energies of intrinsic defects in GaSb are shown in Fig.~\ref{fig-gform} as a function of $E_F$, referenced to the VBM, for Sb-poor and Sb-rich conditions. Ga$_{\mathrm{Sb}}$ dominates in Sb-poor conditions; it has a formation energy under 1 eV and is negatively charged for all values of $E_F$ within the band gap, with an adiabatic transition from the $-$ to $2-$ state, $(-/2-)$, occurring at $E_F=0.16$ eV above the VBM. Such a low energy, negatively charged defect indicates an intrinsically $p$-type material, as is observed experimentally.~\cite{gasb_expt_ptype_hall_physrev_leifer1954,gasb_expt_crystal_growth_hole_conc_jphyschemsol_vandermeulen1967,gasb_expt_ptype_undoped_semicondscitech_haywood1988,gasb_expt_ptype_acceptors_apl_kala2015,gasb_expt_pdoped_effmass_qw_prb_karalic2019} All other defects have formation energies of at least 1 eV higher than Ga$_{\mathrm{Sb}}$ for $E_F$ within the band gap. Previous calculations by Hakala  \textit{et al.}, using DFT-LDA,~\cite{gasb_lda_intrinsic_defects_jap_hakala2002} and Virkkala \textit{et al.},~\cite{gasb_dft_hybrid_intrinsic_defects_prb_virkkala2012} using hybrid DFT, both found that Ga$_{\mathrm{Sb}}$ had the lowest formation energy for $E_F$ in the upper half of the band gap, but predicted compensation by Ga interstitials (Ga$_i^+$), resulting in an insulating material. The LDA calculations did not include the SOI nor any correction for the band gap underestimation, while the hybrid DFT calculations did not include the SOI and used higher convergence criteria than those we employ;~\cite{gasb_dft_hybrid_intrinsic_defects_prb_virkkala2012} their results contradict the experimentally observed $p$-type activity of undoped GaSb.

In Sb-rich conditions, we find that $E_f(\mathrm{Ga}_{\mathrm{Sb}})$ increases significantly, while $E_f(V_{\mathrm{Ga}})$ and $E_f(\mathrm{Sb}_{\mathrm{Ga}})$ both decrease, so that the lowest energy defects are Sb$_{\mathrm{Ga}}$ for $E_F<0.36$ eV and $V_{\mathrm{Ga}}$ for $E_F>0.42$ eV, with Ga$_{\mathrm{Sb}}$ having the lowest energy for $E_F$ between these ranges. As Sb$_{\mathrm{Ga}}$ are positively charged and Ga$_{\mathrm{Sb}}$ and $V_{\mathrm{Ga}}$ negatively charged for $E_F$ within the band gap, these defects self compensate and one would expect $E_F$ to remain trapped roughly mid-gap, resulting in an intrinsically insulating material (we note that the formation energy of Ga$_i$ is also low in this range of $E_F$ and we expect that this defect will play a minor r\^{o}le in the self-compensation mechanism). These formation energies suggest significant concentrations of $V_{\mathrm{Ga}}$ will be present, in agreement with PAS studies,~\cite{gasb_expt_positron_hall_shallow_acceptor_apl_ling2004,gasb_expt_positron_point-defects_jap_kujala2014,gasbn_expt_dft_positron_defects_jap_segercrantz2015,gasb_expt_positron_vsb_instable_prb_segercrantz2017} but the insulating nature contradicts the $p$-type activity of GaSb observed in many differently produced samples. It may be the case that, in non-equilibrium growth techniques, formation of the compensating Sb$_{\mathrm{Ga}}$ may be suppressed, which would result in a $p$-type material where the hole concentration arises from the ionisation of $V_{\mathrm{Ga}}$ and Ga$_{\mathrm{Sb}}$.~\cite{gasb_expt_sdoped_mbe_compensation_jap_lee1990,gasb_expt_tedoped_highmob_jvacscitechb_turner1993} Our results for Sb-rich conditions agree qualitatively with those of Virkkala \textit{et al.},~\cite{gasb_dft_hybrid_intrinsic_defects_prb_virkkala2012} although they did not predict that the $V_{\mathrm{Ga}}$ would become the lowest energy defect for any value of $E_F$ within the band gap. Comparisons with the LDA calculations of Hakala \textit{et al.}~\cite{gasb_lda_intrinsic_defects_jap_hakala2002} are more difficult, as they only reported formation energies for Sb$_{\mathrm{Ga}}$ in the neutral state. We note, however, that they also found $V_{\mathrm{Ga}}$ to be the lowest energy defect close to the conduction band minimum (CBM).

\begin{figure}[ht!]
\centering
\vspace{0.5cm}
\includegraphics*[width=1.0\linewidth]{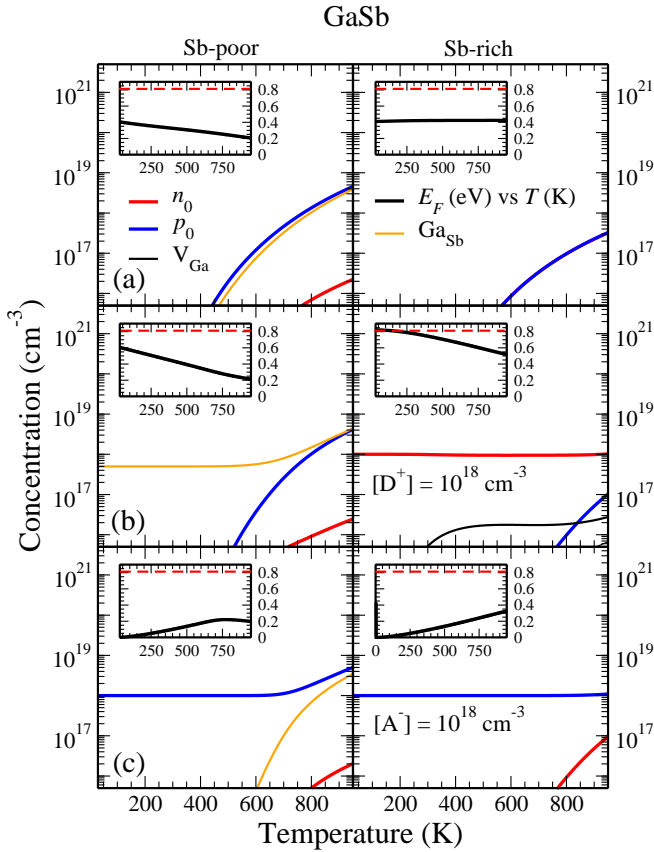}
\caption{(Color online) Concentrations of electron ($n_0$) and hole ($p_0$) carriers and defects (vacancies, interstitials and antisites; see text for description) in GaSb as a function of temperature $T$ calculated for (a) equilibrium conditions, (b) in the presence of a fixed concentration of donors $[$D$^+]=10^{18}$ cm$^{-3}$ and (c) a fixed concentration of acceptors $[$A$^-]=10^{18}$ cm$^{-3}$. The results are shown for Sb-poor and Sb-rich conditions in the left- and right-side panels, respectively. The insets show the self-consistent Fermi energy $E_F$ as a function of $T$, with the conduction band minimum indicated by the red dashed line.}
\label{fig-gconc}
\end{figure}

From our computed defect formation energies and total DOS, we have calculated the self-consistent $E_F$ and equilibrium carrier and defect concentrations by applying the constraint of overall charge neutrality to our system. The results are shown in Fig.~\ref{fig-gconc}(a) over the $T$ range below the melting point (985 K~\cite{crc_handbook_89th_2008}). It is worth noting here that, when varying $T$ in this analysis and for the case of InSb below we do not take into account the variation in band gap, which can be substantial for these narrow gap semiconductors. Indeed, at room temperature the band gap reduces by 86 meV for GaSb~\cite{gasb_expt_review_jap_dutta1997} and 67 meV for InSb,~\cite{insb_expt_bandgap_t_apl_litter1985} compared with their extrapolated 0 K values. Such reductions are a result of thermal expansion and increased electron-phonon coupling, the modelling of which is beyond the scope of this study on defects in both systems. Including the experimental variation in $E_g$ with $T$ in our calculations is not straightforward, as the defect transition levels vary with $T$ in a non-trivial manner. If we do include just the experimental $E_g$ variation, we calculate slightly different electron and hole concentrations which do not alter our conclusions significantly. As modelling temperature effects on the defect formation and transition levels is beyond the scope of the current work, we present our analysis below with the band gap fixed for all temperatures studied. We expect that, at higher $T$, where the band gap is reduced and consequently the electron and hole concentrations increased, compensating defect formation energies will also be lowered as vibrational entropy contributions to the free energy become more significant, so that the changes in concentrations will approximately cancel each other.

From our analysis we find that, in Sb-poor conditions, GaSb is $p$-type with hole concentrations $p_0$ of $\sim 10^{16}-10^{18}$ cm$^{-3}$ for $400<T<800$ K. The source of the $p_0$ is the formation and ionisation of Ga$_{\mathrm{Sb}}$; $p_0$ is equal to $2[\mathrm{Ga}_{\mathrm{Sb}}]$, which is consistent with the dominant charge state of Ga$_{\mathrm{Sb}}$ being $2-$, but at $T\approx 800$ K the concentrations become close to being equal, as $E_F$ moves closer to the VBM where the $-$ state dominates. These calculated hole concentrations are lower by about an order of magnitude than those seen in experiment;~\cite{gasb_expt_crystal_growth_hole_conc_jphyschemsol_vandermeulen1967,gasb_expt_ptype_undoped_semicondscitech_haywood1988} the discrepancy may be due to unwanted impurities such as C that can be introduced during experimental growth, which are not accounted for here. $p_0$ and $[\mathrm{Ga}_{\mathrm{Sb}}]$ are also about an order magnitude lower than those computed by Hakala \textit{et al.},~\cite{gasb_lda_intrinsic_defects_jap_hakala2002} which can be attributed to their lower value of $E_f(\mathrm{Ga}_{\mathrm{Sb}}^{2-})$. The difference in formation energies is probably due to a combination of the difference in functional and in the more crude image charge corrections used in their much earlier work. In Sb-rich conditions, we find that $E_F$ remains trapped at about 0.4 eV above the VBM over the range of $T$ investigated, due to the self-compensating defect physics, whereby the combined concentration of $\mathrm{Sb}_{\mathrm{Ga}}^+$, $\mathrm{Sb}_{\mathrm{Ga}}^{2+}$ and $\mathrm{Ga}_i^+$ equals that of $V_{\mathrm{Ga}}^-$, $V_{\mathrm{Ga}}^{2-}$ and $\mathrm{Ga}_{\mathrm{Sb}}^{2-}$, with the individual proportions depending on $T$. Consequently, the electron concentration $n_0$ is equal to $p_0$ and the material is intrinsically insulating. This insulating nature is rarely seen experimentally; again, unwanted $p$-type impurities not included in this study, as well as non-equilibrium defect formation, expected to be important in samples grown epitaxially where kinetics dominate,~\cite{gasb_expt_ptype_undoped_semicondscitech_haywood1988,gasb_expt_pdoped_effmass_qw_prb_karalic2019} may account for the discrepancy.

When imposing the charge neutrality constraint to determine the self-consistent $E_F$, it is possible to introduce fixed concentrations of other charged defects and calculate the equilibrium carrier and intrinsic defect concentrations in their presence. In this way, one can analyse compensation of fully ionised impurities in an approximate manner. By assuming a fixed concentration of some ionised donor, $[$D$^+]=10^{18}$ cm$^{-3}$, we have calculated donor compensation in GaSb, with our results shown in Fig.~\ref{fig-gconc}(b). We find that, in Sb-poor conditions, rather than introducing $n$-type carriers, the donors are compensated by Ga$_{\mathrm{Sb}}^{2-}$, so that $[$D$^+]=2[\mathrm{Ga}_{\mathrm{Sb}}]$ for $T<600$ K. We see, therefore, that in Sb-poor conditions donor doping will not be effective, assuming that defect formation occurs in equilibrium. In fact, $p_0$ will become greater than $10^{16}$ cm$^{-3}$ at about $T=600$ K, and continues to rise with temperature as $[\mathrm{Ga}_{\mathrm{Sb}}]$ increases above the value necessary to compensate $[$D$^+]$ due to thermal activation, while $E_F$ is pushed closer to the VBM. In Sb-rich conditions, however, we have $[$D$^+]=n_0$ for most of the temperature range studied, so that GaSb will be doped effectively. At lower temperature, $E_F$ remains close to the CBM, but decreases into the band gap with increasing temperature. There is a very small dip in $n_0$ around $T=400$ K, which occurs as thermally induced concentrations of $V_{\mathrm{Ga}}$ compensate slightly the donors. We note that, in MBE-grown samples intentionally doped $n$-type, increasing the V/III ratio (i.e. going towards increasingly Sb-rich conditions) caused a slight increase in compensating acceptor concentrations,~\cite{gasb_expt_sdoped_mbe_compensation_jap_lee1990,gasb_expt_tedoped_highmob_jvacscitechb_turner1993} contrary to our findings here. The effect is small and may be due to non-equilibrium defect formation and/or the presence of unwanted impurities.

In the same way, we can analyse acceptor compensation in GaSb. In Fig.~\ref{fig-gconc}(c), we show the equilibrium carrier and intrinsic defect concentrations in the presence of a fixed concentration of an ionised acceptor, $[$A$^-]=10^{18}$ cm$^{-3}$. The situation here is quite different to donor compensation discussed above; in both Sb-poor and Sb-rich conditions the acceptors are uncompensated and we have a $p$-type material with $p_0=[$A$^-]$. $E_F$ remains close to the VBM, but moves towards mid-gap as $T$ increases, as one would expect due to $T$-induced intrinsic carrier generation. In Sb-poor conditions, for $T>600$ K, substantial concentrations of Ga$_{\mathrm{Sb}}$ form, which further contribute to the $p$-type activity. We therefore find that GaSb can be effectively $p$-doped, whether in Sb-rich or Sb-poor conditions, a result that is consistent with experiment.

\subsection{Defects in InSb}
\label{insb}

\begin{figure}[ht!]
\centering
\vspace{0.5cm}
\includegraphics*[width=1.0\linewidth]{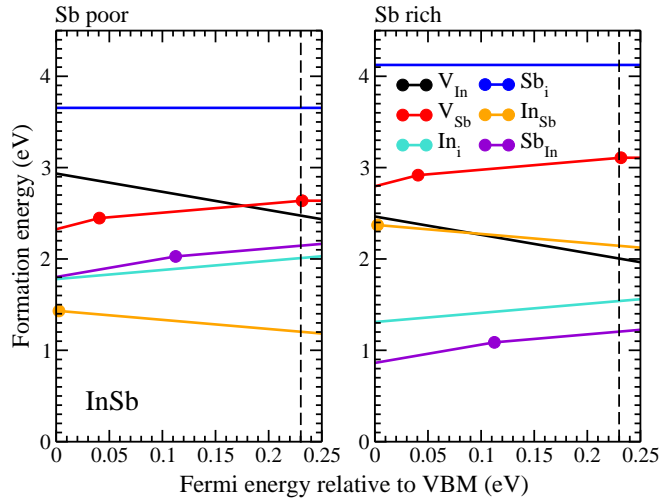}
\caption{(Color online) Calculated formation energies of each intrinsic defect (vacancies, interstitials and antisites; see text for description) in InSb as a function of Fermi energy relative to the valence band maximum (VBM), shown for Sb-poor and Sb-rich conditions. The slope of each line indicates the defect charge state; the transition levels lie where the slopes change. The dashed line indicates the position of the conduction band minimum.}
\label{fig-iform}
\end{figure}

We show our calculated intrinsic defect formation energies as a function of $E_F$ referenced to the VBM in Fig.~\ref{fig-iform}. We find that, in contrast to the case of GaSb, we have a positively charged defect, Sb$_{\mathrm{In}}$, dominating in Sb-rich conditions and a negatively charged defect, In$_{\mathrm{Sb}}$ ,dominating in Sb-poor conditions. Consequently, one would expect an $n$-type material if grown in Sb-rich conditions, and a (weakly, due to the relatively high formation energy) $p$-type material if grown in Sb-poor conditions. Experimentally, both $n$- and $p$-type unintentionally doped samples are routinely prepared, and InSb can be doped relatively easily with electrons or holes as majority carriers.~\cite{insb_expt_hall_conductivity_effmass_physrev_hrostowski1955,insb_expt_fermi_energy_t_pssb_zukotynski1970,insb_expt_intrinsic_carrierconc_jap_chen1972,insb_expt_carrierconc_effmass_vs_t_hall_jphyschemsol_oszwaldowski1988,insb_expt_gasb_defects_compensate_intjhighspeedelecsys_pino2004,insb_expt_growth_sb_antisite_jcrystgrowth_jin2011} Hoglund \textit{et al.}~\cite{insb_dft_inp_inas_lda_bulk_110_defects_prb_hoglund2006} calculated the defect formation energies using DFT-LDA, finding results consistent with ours for Sb-rich conditions, but for the Sb-poor conditions they found that In$_i$ would dominate, resulting in an $n$-type material, in contrast to our results. In their calculations, they found InSb to be gapless, contradicting experiment, and did not discuss corrections for this error nor for image charge interactions in their supercell model. The Sb$_{\mathrm{In}}$ defect has been proposed to be a source of intrinsic $n$-type carriers in epitaxially grown InSb, but can be removed effectively by decreasing the V/III ratio, i.e. moving away from Sb-rich conditions.~\cite{insb_expt_growth_sb_antisite_jcrystgrowth_jin2011} Such an observation is consistent with our calculated formation energies. Vacancies have also been proposed to be important in InSb,~\cite{insb_expt_selfdiffusion_jap_kendall1969,insb_expt_mbe_growth_chinphyslett_zhao2017,insb_expt_vacancies_highT_crysrestech_morozov1986,insb_expt_thermoelectric_zt1_jmaterchema_xin2018} but our results show that their concentrations should be small as their formation energies are relatively high. We note that, although we have pointed out some differences between the defect physics of InSb and GaSb, some of these differences can be traced to the much lower band gap of InSb, compared with GaSb (0.23 eV \textit{vs} 0.808 eV). Restricting the range of $E_F$ to remain less than 0.23 eV in GaSb would result in a similar transition level diagram to that of InSb. This result indicates a small valence band offset between the materials, consistent with earlier studies.~\cite{ii-v_ii-vi_expt_valence_bands_xps_prb_ley1974,insb_gasb_calc_bandoffset_prb_magri2002,iii-v_review_bandfeatures_jap_vurgaftman2001}

\begin{figure}[ht!]
\centering
\vspace{0.5cm}
\includegraphics*[width=1.0\linewidth]{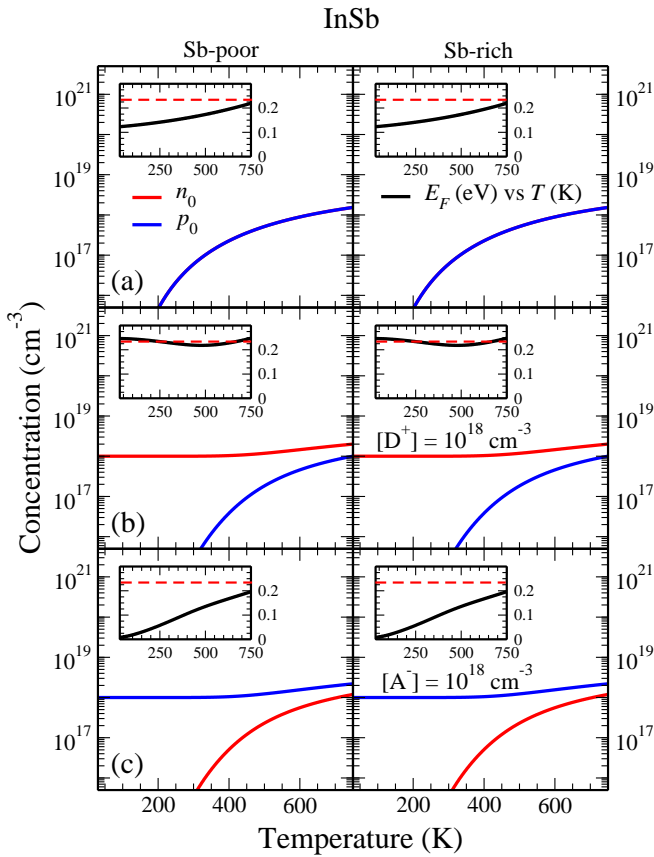}
\caption{(Color online) Concentrations of electron ($n_0$) and hole ($p_0$) carriers and, defects (vacancies, interstitials and antisites; see text for description) in InSb as a function of temperature $T$ calculated for (a) equilibrium conditions, (b) in the presence of a fixed concentration of donors $[$D$^+]=10^{18}$ cm$^{-3}$ and (c) a fixed concentration of acceptors $[$A$^-]=10^{18}$ cm$^{-3}$. The results are shown for Sb-poor and Sb-rich conditions in the left- and right-side panels, respectively. The insets show the self-consistent Fermi energy $E_F$ as a function of $T$, with the conduction band minimum indicated by the red dashed line.}
\label{fig-iconc}
\end{figure}

As with the case of GaSb, we have calculated equilibrium carrier and defect concentrations in InSb (excluding the variation in $E_g$ with $T$, see the discussion above); our results are shown in Fig.~\ref{fig-iconc}(a) over the $T$ range below the melting point (797 K~\cite{crc_handbook_89th_2008}). Despite the dominance of positively and negatively charged defects in Sb-rich and Sb-poor conditions respectively, we find that, under either condition InSb will be insulating as-grown. This result is a consequence of the low band gap and relatively high defect formation energies; thermally induced intrinsic carrier formation will dominate as defect concentrations remain several orders of magnitude below the carrier concentrations over the relevant $T$ range (in Sb-poor conditions, $[\mathrm{In}_{\mathrm{Sb}}]$, not shown in the figure, rises above $10^{14}$ cm$^{-3}$ only for $T>700$ K). $E_F$ remains closer to the CBM, as the DOS at the bottom of the conduction band is much lower than that at the top of the valence band. To produce $n$- and $p$-type samples therefore, one needs to dope the material and nominally undoped samples that have substantial carrier concentrations probably have unwanted impurities present, according to our results. 

In Fig.~\ref{fig-iconc}(b) we show the equilibrium carrier and defect concentrations in the presence of a fixed concentration of ionised donors, $[$D$^+]=10^{18}$ cm$^{-3}$. In both Sb-poor and Sb-rich conditions, we find that InSb can be donor doped effectively, resulting in $n_0=[$D$^+]$ for much of the $T$ range. As the DOS is relatively low at the CBM, to induce the relevant electron concentration $E_F$ is pushed very up to the CBM (see the inset in Fig.~\ref{fig-iconc}(b)). No significant defect compensation is observed; indeed, we find that, for $T>400$ K, thermal ionisation increases $n_0$ above $[$D$^+]$.

We have also analysed acceptor compensation in InSb by assuming a fixed ionised acceptor concentration, $[$A$^-]=10^{18}$ cm$^{-3}$, and computing the resultant carrier and defect concentrations; our results are shown in Fig.~\ref{fig-iconc}(c). In both Sb-poor and Sb-rich conditions there is no effective compensation of the acceptors by defects, indicating that InSb will be easily acceptor doped in either extreme condition. $E_F$ varies across the gap as $T$ increases, which induces minority carrier concentrations while also increasing the majority carrier concentration. We therefore see that InSb can be both $n$- and $p$-doped without significant compensation by intrinsic point defect formation, a result that is consistent with experiment.~\cite{insb_expt_gasb_defects_compensate_intjhighspeedelecsys_pino2004,insb_dft_inp_inas_lda_bulk_110_defects_prb_hoglund2006,insb_expt_growth_sb_antisite_jcrystgrowth_jin2011}

\section{Summary}
\label{conclude}

We have investigated the intrinsic defect physics in GaSb and InSb by computing native defect formation energies using hybrid DFT. We justify our approach by first calculating a range of bulk properties of both systems, obtaining results in good agreement with experiment. We find that, in GaSb Ga$_{\mathrm{Sb}}$ will dominate in Sb-poor conditions, resulting in a $p$-type material, while in Sb-rich conditions self-compensation will occur and the material will be intrinsic. We confirm these inferences from the formation energy calculations by computing equilibrium carrier and defect concentrations as a function of temperature, then study donor and acceptor compensation by assuming fixed concentrations of ionised dopants. We find that GaSb can be easily $p$-doped, but in equilibrium conditions, should only be effectively $n$-doped under Sb-rich conditions. For InSb, we find that positively charged (Sb$_{\mathrm{In}}$) and negatively charged antisite defects (In$_{\mathrm{Sb}}$) dominate in Sb-rich and Sb-poor conditions, respectively. By calculating equilibrium carrier and defect concentrations, however, we show that the material will be intrinsic as-grown, due to the relatively high formation energies, low band gap and consequent thermally induced carrier generation. As the concentrations of compensating defects remain low over the relevant $T$ range, InSb can be effectively $n$- and $p$-doped. Our study provides crucial information on the defect physics of GaSb and InSb, important semiconductors for a range of technological applications.

\section*{Acknowledgment}

The authors acknowledge funding from EPSRC grants ED/D504872, EP/K016288/1 and EP/I01330X/1 and the European Research Council (grant 758345). The authors also acknowledge the use of the UCL Legion and Grace High Performance Computing Facilities (Legion@UCL and Grace@UCL) and associated support services, the IRIDIS cluster provided by the EPSRC funded Centre for Innovation (EP/K000144/1 and EP/K000136/1), the Thomas supercomputer via the U.K. Materials and Modelling Hub (EPSRC grant EP/P020194/1) and the ARCHER supercomputer through membership of the UK's HPC Materials Chemistry Consortium, which is funded by EPSRC grants EP/L000202 and EP/R029431, in the completion of this work. D. O. S. and T. D. V. acknowledge membership of the Materials Design Network.

\bibliographystyle{aip} 


\end{document}